\documentclass[12pt]{iopart}
\usepackage{graphicx}
\usepackage{amssymb}
\begin{document}

\title[Gapless modes of vortices in superfluid $^3$He-B]{Low energy dynamics of gapless and quasi-gapless modes of vortices in superfluid $^3$He-B}
\author{A J Peterson$^{1}$ and M Shifman$^{1,2}$}

\address{$^{1}$ School of Physics and Astronomy, University of Minnesota, Minneapolis, MN 55455, USA \\
$^{2}$ William I. Fine Theoretical Physics Institute, 
University of Minnesota, Minneapolis, MN 55455, USA \\}
\ead{pete5997@umn.edu, shifman@umn.edu}

\begin{abstract}
We discuss the low energy effective dynamics of gapless excitations of the mass vortices of systems similar to the Ginzburg-Landau description of superfluid helium-3 in the bulk B phase.  Our approach is to determine the vortex solution by considering a specific ansatz for the order parameter and minimizing the free energy.  The conditions on the $\beta_i$ coefficients required for the stability of the various solutions for the order parameter are calculated.  By considering the symmetries that are broken by the vortex solutions we are able to generate the moduli fields associated with the low energy excitations of the vortices.  Using these fields we determine the effective free energy describing the dynamics of these excitations.
\end{abstract}
\pacs{67.30.he, 11.10.-z, 47.32.C-, 71.70.Ej}
\submitto{\JPCM}
\maketitle

\section{Introduction}

In recent years there has been a growing interest among high energy theorists in condensed matter systems with non-Abelian group structure.  Such condensed matter systems share a deep connection with many models of high energy physics due to the universality of effective field theories describing the low energy dynamics of the underlying microscopic physics  \cite{Volovik:2006a,Babaev:2001zy}.  Because of this universality, studying condensed matter systems can offer insight on many of the unsolved problems in high energy particle physics and cosmology, and vice versa.

In particular condensed matter systems such as superfluid helium-3 as well as other systems with similar symmetry structure have attracted much attention from the high energy community.  The non-Abelian group structure and the tensorial nature of the order parameter results in a wealth of phenomena that share close similarities with high energy theories, specifically Yang-Mills theories \cite{Volovik:2006a}.  The degeneracy of the order parameter in the B-phase ground state allows for the existence of topologically stable mass vortices of the $\mathbb{Z}$-type similar to flux tubes presenting string-like solitons in four-dimensional Yang-Mills theories \cite{Gorsky:2004ad, Hanany:2003hp, Auzzi:2003fs, Shifman:2004dr, Hanany:2004ea,Eto:2005yh}.  The Nambu-Goldstone excitations of these vortices in $^3$He-B can be described by a low energy effective field theory using the same methods as those used to determine the low energy dynamics of the flux tubes in Yang-Mills theories \cite{Nitta:2013mj}.  The Kelvin modes and $U(1)$ axial rotational modes of $^3$He-B vortices have been thoroughly investigated both theoretically and experimentally  \cite{Thomson:1880, Sonin:1987zz, Simula:2008a, Fonda,Krusius:1984a, Pekola:1985a, Hakonen:1983a}, however it has been recently shown that additional non-Abelian orientational modes may emerge in certain cases \cite{Nitta:2013mj}.  Specifically, the cases we are referring to are those systems with a Ginzburg-Landau free energy of the form of superfluid $^3$He with small parameters $\gamma_2$ and $\gamma_3$ (see (\ref{GLFE}) below).  This requirement is imposed to ensure that the non-Abelian modes from $SO(3)_{S+L}$ are independent or nearly independent of the modes associated with coordinate rotations from $SO(3)_{\rm coord}$.  An analogous case can be discussed in the context of elasticity in two dimensions where the unphysical case of vanishing bulk modulus leads to an enhanced symmetry under rotations $O(2) \rightarrow O(2) \times O(2)$.  Such an enhanced symmetry plays a role in the equivalence of scale and conformal symmetry in the theory of elasticity \cite{Riva:2005gd}.  If we were to consider $\gamma_2,\gamma_3 \sim \gamma_1$ the non-Abelian modes would no longer be quasi-gapless as we will show below.  

The case of $\gamma_2,\gamma_3 \rightarrow 0$ could be approximately achieved in an ultra-cold fermi gas with p-wave pairing, but strictly speaking, superfluid $^3$He does not satisfy this condition.  All of our assertions below as well as any mention of Ginzburg-Landau description of superfluid $^3$He will refer to the case where $\gamma_2,\gamma_3$ are either vanishing or small relative to $\gamma_1$.  In these cases the non-Abelian modes are either gapless or quasi-gapless respectively.

Non-Abelian orientational modes localized on mass vortices also appear in Yang-Mills theories admitting a $U(1) \times SU(N)$ gauge symmetry \cite{Gorsky:2004ad, Hanany:2003hp, Auzzi:2003fs, Shifman:2004dr, Hanany:2004ea, Eto:2005yh}.  Typically the moduli fields associated with the orientational modes are described by a sigma model at low energy \cite{Shifman:2012zz,Shifman:2013oia}, and it is expected that a similar situation should occur for the non-Abelian modes of $^3$He-B \cite{Nitta:2013mj}.  The appearance of these additional modes is of course due to the tensor type order parameter and the symmetry structure of the Ginzburg-Landau free energy.

The tensorial nature of the order parameter results from the behavior of the $^3$He atoms at low temperatures.  At the critical temperature, the helium atoms condense to form pairs similar to Cooper pairing in the BCS description of superconductivity \cite{Leggett:1972a}.  However, unlike conventional superfluids and superconductors, which are described by a wave function with a single complex component, the pairing in superfluid $^3$He occurs in p-wave $L=1$ orbitals due to the short range repulsive core behavior of the interaction between helium-atoms.  Due to Fermi-Dirac statistics the spin degree of freedom is required to from a symmetric $S=1$ pairing, and thus the complete order parameter is given by a $3 \times 3$ matrix with complex components \cite{Leggett:1975a,Thuneberg:1987a,Sauls:1981}.  Thus, in the absence of spin-orbit and magnetic interactions, the superfluid $^3$He system is described by a Ginzburg-Landau free energy possessing a symmetry group 
\begin{equation}
G = U(1)_P \times SO(3)_S \times SO(3)_L,
\label{SymmetryGroup}
\end{equation}
where $U(1)_P$ is the group of global phase rotations of the order parameter.  In the B-phase ground state the symmetry group $G$ is spontaneously broken down by the order parameter retaining a spin-orbit locked symmetry
\begin{equation}
G \rightarrow H_B = SO(3)_{S+L}.
\label{BGroup}
\end{equation}
This type symmetry breaking also occurs in the theory of color superconductivity observed in quark matter at high density where the color $SU(3)_C$ and flavor $SU(3)_F$ symmetries are spontaneously broken to a diagonal color-flavor locked $SU(3)_{C+F}$ symmetry \cite{Alford:2007xm}.

The existence of topologically stable mass vortices in $^3$He-B results from the $U(1)_P$ phase degeneracy of the order parameter in the B-phase ground state.  The gapless excitations of the vortices are determined by the moduli fields that are generated from symmetry transformations acting nontrivially on a given vortex solution.  The goal of the present work is to explore such vortex excitations using techniques that are standard to high energy theory.  Particularly, we are interested in exploring the non-Abelian gapless and quasi-gapless modes localized on the vortex axis and their interactions with the well known Kelvin modes and $U(1)$ axial modes.  The general framework for determining the low energy dynamics of vortices in systems like superfluid $^3$He-B was outlined in \cite{Nitta:2013mj}, and we will follow the same procedure in our analysis.  Here we will consider specific forms of the order parameter describing vortex solutions, and carry out the calculation of the effective free energy of the gapless and quasi-gapless modes.  In general the order parameter $e_{\mu i}$ is a complex $3 \times 3$ matrix function of the coordinates $\vec{x}_{\perp}=(x,y)$ perpendicular to the vortex axis, which is convenient to decompose into its trace, symmetric, and anti-symmetric components
\begin{eqnarray}
e_{\mu i} &=\frac{1}{3}e_{\sigma \sigma} \, \delta_{\mu i} + e^S_{\mu i} + e^A_{\mu i}, \nonumber \\
e^S_{\mu i} &= e_{\{\mu i\} }-\frac{1}{3}e_{\sigma \sigma}\delta_{\mu i}, \nonumber \\
e^A_{\mu i} &= e_{[\mu i]} \equiv \varepsilon_{\mu i k}\chi^k.
\label{CompleteDecomposition}
\end{eqnarray}
As was suggested in \cite{Nitta:2013mj} we will restrict the order parameter to the trace and anti-symmetric parts
\begin{equation}
e_{\mu i}=e^{i \phi}f(\vec{x}_{\perp})\delta_{\mu i}+\varepsilon_{\mu i k}\chi^k(\vec{x}_{\perp}).
\label{RestrictedDecomposition}
\end{equation}
This restriction is imposed to both simplify the analysis, and contain enough complexity to illustrate the dynamics of the non-Abelian orientational modes of the vortex solution.  However, we hasten to emphasize that this ansatz is an oversimplification from both experimental \cite{Krusius:1984a, Pekola:1985a, Hakonen:1983a} and theoretical standpoints \cite{Thuneberg:1986a, Volovik:1986a}.  More general forms of the order parameter will be considered in future work.

The number of low energy non-Abelian modes will be dependent on the symmetry of the order parameter (\ref{RestrictedDecomposition}).  When $\gamma_2,\gamma_3 = 0$ we will find either two or three additional non-Abelian modes depending on whether the solution retains a $U(1)_z$ axial symmetry or not.  In the case of an axially symmetric solution we will show that there are two moduli living on the degeneracy space given by $SO(3)_{S+L}/U(1) \simeq S^2$.  An axially asymmetric solution has three moduli on the degeneracy space $SO(3)_{S+L}/1 \simeq S^3/\mathbb{Z}_2$.  Here we are referring to an axial transformation from the group $SO(2)_{S+L}$ about the $z$-axis.

We mention that we are considering the $U(1)_P$ phase symmetry of $G$ as a global symmetry and thus we may find an additional axial mode generated from this group.  It is immediately emphasized that this situation is distinct from a $U(1)$ gauge symmetry considered in Yang-Mills theories where the additional mode can be removed by a gauge transformation.

For $\gamma_2,\gamma_3 \neq 0$ but small, some of these modes will acquire small mass gaps, and will be considered as quasi-gapless.  The number of physical excitations may further be reduced after quantization due to the nature of gapless modes in the case of non-relativistic systems where the Goldstone Theorem \cite{Goldstone:1961, Goldstone:1962} is more subtle \cite{Nielsen:1976, Watanabe:2012, Hidaka:2012ym}.  We will not discuss this issue in detail here.  The discussion of quantization will be left for future research.

We will begin in section II with a review of the Ginzburg-Landau description of the superfluid $^3$He B phase and discuss the topological arguments leading to the existence of stable vortices in the bulk.  The specific form of the ansatz (\ref{RestrictedDecomposition}) will be determined and the conditions required on the phenomenological parameters of the free energy to allow the formation of an anti-symmetric component will be calculated in section III.  In section IV, we will discuss the symmetries broken by the vortex solutions and determine the number and type of emerging moduli fields due to the broken symmetries.  In section V, we will formulate the emerging effective free energy describing the dynamics of the non-Abelian modes and their interactions with the Kelvin and axial modes.  We will conclude in section VI with a discussion of our results and their relations to similar phenomena in high energy theory.

\section{The Ginzburg-Landau description of superfluid $^3$He-B}

In this section we will briefly review the Ginzburg-Landau theory describing the superfluid phases of $^3$He.  We will follow the description as given by \cite{Volovik:2006a} and \cite{Mineev:1998, Leggett:2006} where more detailed discussions can be found.  As discussed above the order parameter $e_{\mu i}$ is a complex $3 \times 3$ matrix that transforms under the vector representations of $SO(3)_L$ and $SO(3)_S$
\begin{equation}
e_{\mu i} \rightarrow e^{i \psi}S_{\mu \nu} L_{i j} e_{\nu j},
\label{Transform}
\end{equation}
where $e^{i \psi}$ is an element of the global $U(1)_P$ phase rotations.  We write the most general free energy possessing the complete symmetry $G = U(1)_P \times SO(3)_S \times SO(3)_L$
\begin{eqnarray}
&F_{GL}=F_{time}+F_{grad}+V, \nonumber \\
&F_{time}=i e_{\mu i}\partial_t e^{\star}_{\mu i}, \nonumber \\
&F_{grad}=\gamma_1 \partial_i e_{\mu j} \partial_i e^{\star}_{\mu j}+\gamma_2 \partial_i e_{\mu i} \partial_j e^{\star}_{\mu j}+\gamma_3 \partial_i e_{\mu j} \partial_j e^{\star}_{\mu i}, \nonumber \\
&V=-\alpha e_{\mu i} e^{\star}_{\mu i}+\beta_1 e^{\star}_{\mu i} e^{\star}_{\mu i} e_{\nu j} e_{\nu j}+\beta_2 e^{\star}_{\mu i} e_{\mu i} e^{\star}_{\nu j} e_{\nu j} +\beta_3 e^{\star}_{\mu i} e^{\star}_{\nu i} e_{\mu j} e_{\nu j}, \nonumber \\
&+\beta_4 e^{\star}_{\mu i} e_{\nu i} e^{\star}_{\nu j} e_{\mu j} +\beta_5 e^{\star}_{\mu i} e_{\nu i} e_{\nu j} e^{\star}_{\mu j},
\label{GLFE}
\end{eqnarray}
where the parameters $\gamma_i$, $\alpha$, and $\beta_i$ are phenomenological parameters depending on temperature and pressure that can be determined from BCS-like calculations from the underlying microscopic theory \cite{Sauls:1981}, and may include corrections from strong coupling considerations \cite{Choi:2007a}.  In this paper we will adjust the constants at our will depending on the particular features we wish to illustrate.  

The free energy can be minimized by considering the subgroups of the group $G$.  Two of these subgroups can be realized physically, which are characterized by the A phase $H_A = U(1) \times U(1)$, and the B phase $H_B = SO(3)_{S+L}$.  In the bulk A phase the order parameter takes the form
\begin{equation}
(e_0^A)_{\mu i} = \frac{\Delta}{\sqrt{2}}V_i(\Delta'_{\mu}+i\Delta''_{\mu}),
\end{equation}
where $\vec{V}$ is a unit vector in the direction of the spin, and $\vec{\Delta'}$ and $\vec{\Delta''}$ are mutually orthogonal unit vectors whose cross product $\vec{\Delta'} \times \vec{\Delta''}$ is in the direction of the orbital angular momentum \cite{Leggett:1975a}.  In this work we will consider only the bulk B phase characterized by the order parameter
\begin{equation}
(e_0)_{\mu i}=e^{i \psi} \Delta (R_0)_{\mu i},
\label{BulkB}
\end{equation}
where $(R_0)_{\mu i}$ is a generic element of $SO(3)$.  The gap parameter $\Delta$ can be found by inserting (\ref{BulkB}) into the potential $V$ in (\ref{GLFE}) and minimizing the expression.  The result is
\begin{equation}
\Delta=\frac{\alpha}{6\beta_{12}+2\beta_{345}},
\label{Gap}
\end{equation}
where we are employing a shorthand notation
\begin{eqnarray}
&\gamma_{abc...} = \gamma_a + \gamma_b +\gamma_c + ..., \nonumber \\
&\beta_{abc...} = \beta_a + \beta_b + \beta_c + ...
\end{eqnarray}
We recognize that the order parameter (\ref{BulkB}) is invariant under the simultaneous spin and orbital rotations
\begin{equation}
S_{\mu \nu}L_{i j}(R_0)_{\nu j}=(R_0)_{\mu i},
\end{equation}
when $S=R_0LR_0^T$.  Thus we observe the bulk B phase preserves a locked spin-orbit rotational symmetry $H_B=SO(3)_{S+L}$ \cite{Volovik:2006a,Mineev:1998, Leggett:2006}.  The degeneracy of the ground state allows us to choose $(R_0)_{\mu i} = \delta_{\mu i}$.

For our purposes we will restrict our analysis to the case that $\gamma_2,\gamma_3$ are small compared to $\gamma_1$.  In the case that $\gamma_2,\gamma_3 = 0$ the free energy (\ref{GLFE}) is invariant under the separate orbital $SO(3)_L$ and coordinate $SO(3)_{\rm coord}$ rotations.  Thus the complete symmetry group $G$ is enhanced by an additional set of rotational generators from $SO(3)_{\rm coord}$ and we can treat the orbital and spin degrees of freedom as internal symmetries.  As $\gamma_2,\gamma_3$ are increased to small but non-zero values the symmetry group $G$ is explicitly broken so that orbital rotations from $SO(3)_L$ must be performed with the corresponding coordinate rotations from $SO(3)_{\rm coord}$.  We will show below how this explicit breaking affects the gapless modes of the vortex solutions below.

To discuss topological vortices in the bulk B phase we consider the degeneracy space 
\begin{equation}
G/H_B = U(1)_P \times SO(3).
\label{DegeneracyGroup}
\end{equation}
The topologically stable solutions are determined by the first fundamental group
\begin{equation}
\pi_1(U(1)\times SO(3)) = \mathbb{Z}+\mathbb{Z}_2.
\end{equation}
The $\mathbb{Z}$-sector describes the mass vortices, which are characterized by an integer topological winding $n$ of the phase $e^{i \psi} \sim e^{in\phi}$.  The $\mathbb{Z}_2$-sector describes the spin vortices with winding $\nu \in \{0,1\}$.  We will consider the case $n=1, \; \nu = 0$ below. 

Once the existence of stable topological mass vortices is established, the specific form of the order parameter $e_{\mu i}$ describing the vortex solution must be determined by minimizing the free energy (\ref{GLFE}) with the requirement that the solution have a non-trivial winding $n$.  In principle it is necessary to consider the full form of the order parameter, however to simplify the analysis we will assume the form of the order parameter given by \cite{Nitta:2013mj}
\begin{equation}
e_{\mu i} = e^{i \phi}f(\vec{x}_{\perp})\delta_{\mu i} + \varepsilon_{\mu i k} \chi^k(\vec{x}_{\perp}),
\label{ResDec}
\end{equation}
where $\vec{x}_{\perp}=(x,y)$ is the coordinate vector perpendicular to the vortex axis, and $f$ and $\chi^k$ are to be determined by minimizing (\ref{GLFE}).  We will require that $f(\vec{x}_{\perp})$ be real, however we will not necessarily consider $\chi^k(\vec{x}_{\perp})$ to be real.  The boundary conditions required by $f$ and $\chi^k$ due to the winding and B phase vacuum are given by
\begin{eqnarray}
f(r_{\perp} \rightarrow 0) = 0, \; \; \; f(r_{\perp} \rightarrow \infty) = \Delta \nonumber \\
\chi^k(r_{\perp} \rightarrow \infty) = 0.
\label{BoundaryCond}
\end{eqnarray}
Since $\chi^k$ is not restricted by any winding we do not assume any value for $\chi^k$ as $r_{\perp} \rightarrow 0$.  We will establish in the next section the conditions for which a non-trivial $\chi^k$ field develops in the vortex core.

A more complete ansatz, including symmetric components, as well as solutions more closely related to cases encountered experimentally, will be considered in future projects.

\section{Anti-symmetric tensor structure in the vortex core}

Having restricted our ansatz for the order parameter to the form (\ref{ResDec}) we can determine the requirements on the $\beta_i$ coefficients such that a non-trivial vector field $\chi^k$ develops.  We will proceed in similar fashion as discussed in \cite{Witten:1984eb} in the context of superconducting strings.  An analogous calculation has been recently considered in \cite{Monin:2013kza} for the Abrikosov-Nielsen-Olesen (ANO) string \cite{Abrikosov:1957, Nielsen:1973}.  We begin with the case that $\chi^k=0$ everywhere
\begin{equation}
(e_0)_{\mu i} = e^{i \phi}f(r_{\perp})\delta_{\mu i}.
\label{TraceAnsatz}
\end{equation}
Here we are assuming the function $f$ is a real function of $r_{\perp}=|\vec{x}_{\perp}|$.  We proceed by considering the stability of the solution (\ref{TraceAnsatz}) under small perturbations of the $\chi^k$ field.  We will assume for the moment that the small $\chi^k$ field is also a function of $r_{\perp}$ only and consider the case $\gamma_2,\gamma_3 = 0$.  Thus we are free to set
\begin{equation}
\vec{\chi}=(0,0,\chi(r_{\perp})),
\label{OriginalAnsatz}
\end{equation}
due to the independence of orbital and coordinate rotations.  We determine the form of $f(r_{\perp})$ by inserting the ansatz (\ref{TraceAnsatz}) into the free energy (\ref{GLFE}) and minimizing.  The numerical solution of $f(r_{\perp})$ is shown in Figure 1. 

\begin{figure}[h!]
\centering
\includegraphics{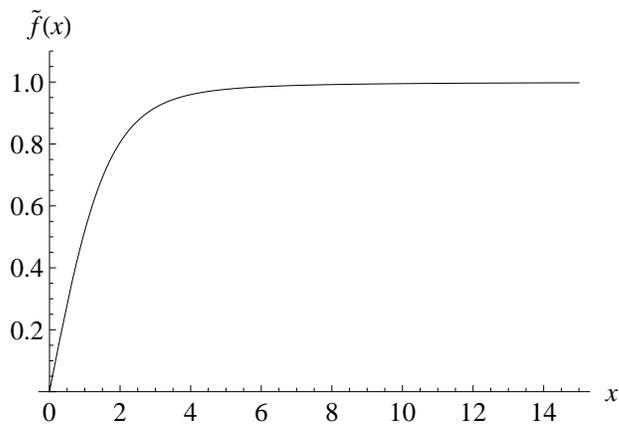}
\caption{The numerical solution $\tilde{f}(x) = f(x)/\Delta$ is plotted.  Here we have defined $x \equiv r_{\perp}\sqrt{\alpha/\gamma_1}$.  For $x \ll 1$ the solution follows the form $\tilde{f}(x) \sim 0.583x$.  In the opposite limit $x \gg 1$ the function $\tilde{f}(x) \rightarrow 1-1/2x^2+\mathcal{O}(x^{-4})$.}
\end{figure}

After determining $f(r_{\perp})$ we now assume a small perturbation $\chi(r_{\perp})$ leaving the form of $f(r_{\perp})$ fixed.  Looking at the $\chi$ dependent part of the free energy (\ref{GLFE}) we find
\begin{equation}
F_{\chi} =\int r_{\perp}dr_{\perp}\left\{ i\chi\frac{\partial \chi}{\partial t}+\chi L_2 \chi +\mathcal{O}(\chi^4) \right\} ,
\label{Fchi}
\end{equation}
where $L_2$ is given given by:
\begin{eqnarray}
&L_2 = -2\gamma_1\frac{1}{r_{\perp}} \frac{\partial}{\partial r_{\perp}} \left(r_{\perp}\frac{\partial}{\partial r_{\perp}}\right)+V(r_{\perp}), \nonumber \\
&V(r_{\perp}) = 4(3\beta_2 + 2\beta_4)f^2(r_{\perp}) -2\alpha.
\label{Schrodinger}
\end{eqnarray}
In this form $L_2$ represents a Schrodinger operator, whose eigenvalues and eigenfunctions can be determined from
\begin{equation}
L_2 \chi_n = \omega_n \chi_n, \mbox{where } \chi=\sum_n a_n \chi_n(r).
\label{EigenValueProb}
\end{equation}
A negative eigenvalue $\omega_0$ of $L_2$ would imply the existence of a solution $\chi_0(r)$ such that $F_{\chi} < 0$, and thus would represent an instability of the ansatz (\ref{TraceAnsatz}).

Solving numerically (\ref{EigenValueProb}) we find that $\omega_0 < 0$ if
\begin{equation}
\frac{1}{2} \le \frac{3\beta_2+2\beta_4}{6\beta_{12}+2\beta_{345}} \lesssim 0.76,
\label{Condition}
\end{equation}
where the lower bound is required to satisfy the boundary condition $\chi(r_{\perp} \rightarrow \infty) \rightarrow 0$.  A numerical solution for $\chi(r_{\perp})$ when this condition is satisfied is shown in Figure 2.

\begin{figure}[h!]
\centering
\includegraphics{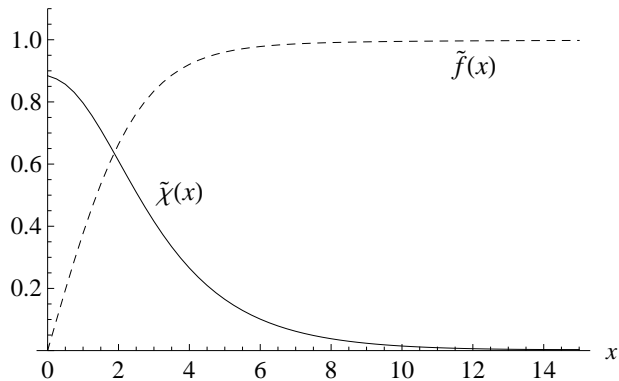}
\caption{The plot shows the numerical solutions for the $\tilde{\chi}(x) \equiv \chi(x)/\Delta$ and $\tilde{f}(x)$, where $x \equiv r_{\perp}\sqrt{\alpha/\gamma_1}$ as previously defined.  The solutions are generated by a numerical minimization of the free energy (\ref{GLFE}), assuming the ansatz (\ref{ResDec}) with the boundary conditions from (\ref{BoundaryCond}).  Note that so far we are working in the limit $\gamma_2,\gamma_3 \rightarrow 0$.}
\end{figure}

We immediately point out that this condition (\ref{Condition}) is not absolute, and represents only the case in which a diagonal ansatz (\ref{TraceAnsatz}) is locally unstable.  If the condition is not satisfied the diagonal ansatz may only be a local minimum and a solution with non-zero $\chi$ may still persist as the absolute minimization of the free energy (\ref{GLFE}).  This is in fact the case both experimentally and theoretically for vortices in superfluid $^3$He-B where the $\beta_i$ coefficients do not satisfy (\ref{Condition}), however the minimizing vortex solutions occur with either an A-phase core, or a double core vortex.  Both solutions break the orbital and coordinate axial symmetry.

The existence of a non-zero $\chi(r_{\perp})$ field would imply the breaking of the $SO(3)_{S+L}$ down to a $U(1)$ rotation, and thus we would expect two non-Abelian moduli from the degeneracy space
\begin{equation}
SO(3)_{S+L}/U(1) \simeq S^2.
\end{equation}
The situation is more interesting if we consider a complex ansatz for the field $\chi^k(\vec{x}_{\perp})$.  In particular we consider the form
\begin{equation}
\vec{\chi}_A(r_{\perp},\phi)=
\left(\begin{array}{c} 
1 \\ 
i \\
0 
\end{array}\right)
\frac{\chi_0(r_{\perp})}{\sqrt{2}}+
\left(\begin{array}{c} 
1 \\ 
-i \\
0 
\end{array}\right)
\frac{\chi_2(r_{\perp})}{\sqrt{2}}e^{2i \phi}.
\label{ComplexAnsatz}
\end{equation}
This ansatz has the form of an A-phase vortex core.  It is easy to show from the equations of motion that
\begin{equation}
\chi_0, \chi_2 \rightarrow \frac{c}{r_{\perp}} \mbox{ as }  r_{\perp} \rightarrow \infty,
\label{BoundaryCondition2}
\end{equation}
where $c$ is a constant that must be determined by solving the equations of motion completely.

Following a similar procedure as above, we find a non-trivial $\chi_0$ field develops if the condition (\ref{Condition}) is satisfied, and thus so does the $\chi_2$ field in order to satisfy (\ref{BoundaryCondition2}).  Figure 3 shows a plot of all three non-trivial fields $f(r_{\perp})$, $\chi_0(r_{\perp})$, and $\chi_2(r_{\perp})$, when the condition (\ref{Condition}) is satisfied.

\begin{figure}[h!]
\centering
\includegraphics{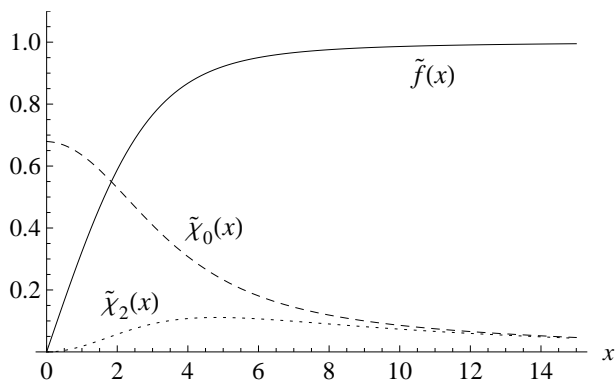}
\caption{The numerical solutions for $\tilde{f}(x) = f(x)/\Delta$, $\tilde{\chi}_0(x) = \chi_0(x)/\sqrt{2}\Delta$, and $\tilde{\chi}_2(x) = \chi_2(x)/\sqrt{2}\Delta$ are plotted, where again $x \equiv r_{\perp}\sqrt{\alpha/\gamma_1}$.  As $x \rightarrow \infty$ we have $\tilde{\chi}_0(x) = \tilde{\chi}_2(x) \rightarrow c/x$, where $c$ is a constant that must be determined by solving completely the vortex solution.}
\end{figure}

In the case of (\ref{ComplexAnsatz}), the non-Abelian symmetry $SO(3)_{S+L}$ is completely broken by the vortex solution and has the degeneracy space
\begin{equation}
SO(3)_{S+L}/1 \simeq S^3/\mathbb{Z}_2.
\end{equation}
This is, of course, due to the fact that the ansatz (\ref{ComplexAnsatz}) is not an eigenvector of a rotation about any axis.  Thus the orbital $U(1)$ symmetry retained by the original ansatz (\ref{OriginalAnsatz}) is now broken by (\ref{ComplexAnsatz}) and we in fact have three non-Abelian moduli.  However, we can see that the solution (\ref{ComplexAnsatz}) still retains a locked $U(1)$ symmetry under simultaneous  orbital, coordinate, and phase rotations,
\begin{equation}
\vec{\chi}(\phi) = e^{i \delta}R_z(\delta)\vec{\chi}(\phi-\delta)R^{-1}_z(\delta),
\label{HiddenSymm}
\end{equation}
where $R_z(\delta)$ is the rotational matrix about the $z$-axis.  Thus, although the orbital $SO(3)_{S+L}$ symmetry is completely broken, one of the additional modes will be equivalent to the $U(1)_A$ axial mode.  

We may further increase the complexity of the ansatz by generalizing (\ref{ComplexAnsatz}) to the following 
\begin{equation}
\vec{\chi}_{dc}(r_{\perp},\phi)=
\left(\begin{array}{c} 
\chi_0(r_{\perp}) \\ 
i\zeta_0(r_{\perp}) \\
0 
\end{array}\right)
\frac{1}{\sqrt{2}}
+
\left(\begin{array}{c} 
\chi_2(r_{\perp}) \\ 
-i\zeta_2(r_{\perp}) \\
0 
\end{array}\right)
\frac{e^{2i \phi}}{\sqrt{2}}.
\label{DoubleCoreAnsatz}
\end{equation}
This ansatz would represent the case of a double core vortex restricted to anti-symmetric off diagonal components.  Again, to satisfy the requirement of cylindrical symmetry and the equations of motion at large distances from the vortex axis we must have
\begin{equation}
\chi_0,\; \chi_2, \; \zeta_0, \; \zeta_2 \rightarrow \frac{c}{r_{\perp}} \mbox{ as } r_{\perp} \rightarrow \infty.
\end{equation}
Figures 4 and 5 show plots of the functions $\chi_0(r_{\perp}), \; \chi_2(r_{\perp}), \; \zeta_0(r_{\perp}),$ and $\zeta_2(r_{\perp})$.

\begin{figure}[h!]
\centering
\includegraphics{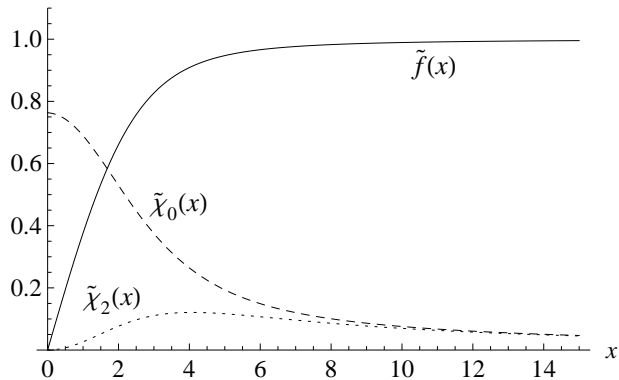}
\caption{The numerical solutions for $\tilde{f}(x) = f(x)/\Delta$, $\tilde{\chi}_0(x) = \chi_0(x)/\sqrt{2}\Delta$, and $\tilde{\chi}_2(x) = \chi_2(x)/\sqrt{2}\Delta$ are plotted, where again $x \equiv r_{\perp}\sqrt{\alpha/\gamma_1}$.}
\end{figure}

\begin{figure}[h!]
\centering
\includegraphics{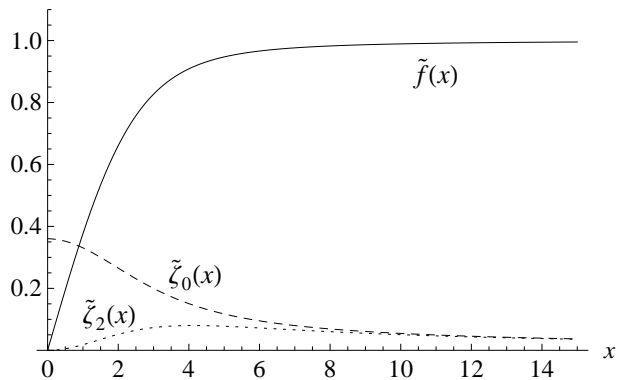}
\caption{The numerical solutions for $\tilde{\zeta}_0(x) = \zeta_0(x)/\sqrt{2}\Delta$, and $\tilde{\zeta}_2(x) = \zeta_2(x)/\sqrt{2}\Delta$ are plotted.}
\end{figure}

Generally speaking a double core vortex would have four moduli from the complete breaking of $SO(3)_{S+L} \times U(1)_A$ since no hidden symmetry is necessarily preserved for arbitrary functions $\chi_0(r_{\perp}), \; \chi_2(r_{\perp}), \; \zeta_0(r_{\perp}),$ and $\zeta_2(r_{\perp})$.

\section{Low energy excitations and Counting Gapless Modes}

In this section, we will discuss the low energy excitations of vortices in the B phase from a general point of view.  We will determine the moduli fields and their interactions by considering the broken symmetries of the free energy (\ref{GLFE}).  In the following section, we will write out the effective free energy for the gapless and quasi-gapless modes for the specific cases (\ref{ResDec}), (\ref{OriginalAnsatz}), (\ref{ComplexAnsatz}), and (\ref{DoubleCoreAnsatz}) considered in the previous section.

Generally the moduli fields of the vortex are determined by the continuous symmetries of the vacuum that are broken by the vortex solution.  We consider first the case that $\gamma_2,\gamma_3 = 0$, and thus we have the addition coordinate rotational $SO(3)_{\rm coord}$ symmetry.  Thus the continuous symmetries of (\ref{GLFE}) including the coordinate rotations $SO(3)_{\rm coord}$ and translations $t$ are given by the group
\begin{equation}
\mathcal{G}=U(1)_P \times SO(3)_S \times SO(3)_L \times SO(3)_{\rm coord} \times t.
\end{equation}
We denote the elements of this group in the following way \cite{Salomaa:1985}
\begin{eqnarray}
&U_{\theta}=e^{i\hat{I}\theta}, \;\hat{I}e_{\mu i} = e_{\mu i}, \; \hat{I}e_{\mu i}^{\star}=-e_{\mu i}^{\star}, \nonumber \\
&\hat{S}_{\beta}e_{\mu i}=-i\varepsilon_{\beta \alpha \nu}e_{\nu i}, \nonumber \\
&\hat{L}^{int}_je_{\mu i} = -i\varepsilon_{jik}e_{\mu k}, \nonumber \\
&\hat{L}^{ext}_je_{\mu i} = -i\varepsilon_{jlk}x_l\frac{\partial}{\partial x_k}e_{\mu i}, \nonumber \\
&t_{\vec{x_0}} e_{\mu i}(\vec{x})=e_{\mu i}(\vec{x} - \vec{x_0}),
\end{eqnarray}
where $U_{\theta} \in U(1)_P$.  We distinguish the orbital and coordinate rotation generators by $\hat{L}^{ext}$ and $\hat{L}^{int}$ respectively.  

The moduli fields are determined from the broken generators of $\mathcal{G}$ which leave the order parameter far away from the vortex core invariant.  We already know that in the vacuum far from the vortex core the order parameter takes the form
\begin{equation}
e_{\mu i} \rightarrow e^{i \phi}\Delta \delta_{\mu i}, \mbox{ as } r \rightarrow \infty.
\label{VortexAsymptote}
\end{equation}
The asymptotic form (\ref{VortexAsymptote}) is invariant under the symmetry group $H_B = SO(3)_{S+L}$ as well as translations $t$ and coordinate rotations about the $x$ and $y$ axes.  The solution is not invariant under coordinate rotations about the $z$-axis, however it is invariant under a simultaneous coordinate rotation about the $z$-axis and a corresponding phase rotation.  The element of this transformation is given by
\begin{equation}
U_{\phi}=e^{i\hat{Q}\phi}, \; \hat{Q}=\hat{L}_z^{ext}-\hat{I}, \; \hat{L}_z^{ext}=-i\frac{\partial}{\partial \phi}.
\label{AxialTrans}
\end{equation}
We denote the group of the transformations (\ref{AxialTrans}) as $U(1)_A$.  Solutions invariant under $U(1)_A$ are known as axially symmetric, although they are strictly speaking not axially symmetric because of the phase winding $e^{i\phi} \in U(1)_P$.

Thus we have three generators $\omega_{x,y,z}$ from $SO(3)_{S+L}$, one generator $\delta$ from the axial symmetry $U(1)_A$, two generators $\hat{L}_{x,y}^{ext}$ from the remaining coordinate rotations of the degeneracy space $SO(3)_{\rm coord}/SO(2)_z$, and two from translations in the $x$ and $y$ directions.  We exclude translations in the $z$ direction as we will assume the vortex solution is independent of $z$ everywhere.  In addition, it is a simple matter to show that coordinate rotations about the $x$ and $y$ axes can be written as $z$ dependent translations in the $xy$-plane \cite{Ivanov:1975, Clark:2003, Low:2001bw, Nitta:2013mj}.  We will denote the two generators of translations as $\xi_{x,y}$.  Thus, we actually have a total of six possible gapless moduli of the vortex solution associated with spin-orbit rotations, axial rotations from $U(1)_A$, and translations:  
\begin{equation}
\omega_x, \; \omega_y, \; \omega_z, \;  \xi_x, \; \xi_y, \mbox{ and } \delta.
\label{ModuliList}
\end{equation}
The particular moduli that appear in the effective theory depend on which of these generators act non-trivially on the vortex solutions (\ref{OriginalAnsatz}), (\ref{ComplexAnsatz}), and (\ref{DoubleCoreAnsatz}).  Table I includes a summary of the type of moduli, total number of moduli, and the degeneracy space of the vortex forms (\ref{TraceAnsatz}), (\ref{OriginalAnsatz}), (\ref{ComplexAnsatz}), and (\ref{DoubleCoreAnsatz}) we considered in the previous section.

When $\gamma_2,\gamma_3 \neq 0$ the $SO(3)_L \times SO(3)_{\rm coord}$ symmetry is explicitly broken to $SO(3)_{L+{\rm coord.}}$, which we will simply write as $SO(3)_L$.  The generator of coordinate rotations is thus
\begin{equation}
\hat{L}_k = \hat{L}_k^{\rm int}+\hat{L}_k^{\rm ext}.
\end{equation}
Thus the orbital rotations about the $x$ and $y$ axes from $SO(3)_{S+L}$ are correlated with the corresponding coordinate rotations.  The $SO(3)_{S+L}$ rotations about the $x$ and $y$ axes are also correlated with the translations in the $xy$-plane due to the local equivalence of the coordinate rotations and translations.  Only axial transformations about the $z$-axis from $U(1)_A$ are distinct.  We have only three available gapless moduli fields in this case:
\begin{equation}
\xi_x, \; \xi_y, \mbox{ and } \delta.
\label{RefinedModuliList}
\end{equation}

When $\gamma_2,\gamma_3$ are small but non-zero we may include all six moduli fields from (\ref{ModuliList}).  However as we will see below some of the moduli will acquire a small mass gap proportional to $\gamma_{23}$.  In addition, we will find that the moduli $\omega_{x,y}$ will acquire interactions with $\xi_{x,y}$ due to the equivalence of coordinate rotations and translations.  We will call the moduli with a small mass gap, quasi-gapless.

\begin{table}
\caption{\label{label}A summary of degeneracy space and associated moduli for the vortex solutions when $\gamma_2,\gamma_3 =0$ considered in the previous section is shown.  The first column indicates the core type solution.  Columns 2-5 indicate the moduli fields emerging in the various solutions.  The degeneracy spaces in the sixth column are denoted with subscripts indicating the group associated with the degeneracy.  Additionally, we have defined $J \equiv S+L$.  The last column shows the total number of emerging moduli.}
\begin{indented}
\item[]\begin{tabular}{@{}lllllcc}
\br
Core Type & $\vec{\xi}_{\perp}$ & $\omega_{x,y}$ & $\omega_z$ & $\delta$ & Degeneracy Space  & Number of Moduli \\ 
\mr
$\vec{\chi}=0$ & $\checkmark$ & $\times$ & $\times$ & $\times$ & $S^2_t$ & 2\\
$\vec{\chi}=\chi_z(r_{\perp})\hat{z}$ & $\checkmark$  &  $\checkmark$  & $\times$ & $\checkmark$ & $S^2_{J}\times S^1_A \times S^2_t$ & 5\\ 
A-phase & $\checkmark$ & $\checkmark$ &  \multicolumn{2}{c}{$\omega_z \sim \delta^{\rm a}$} & $S^2_{J+A} \times S^1_{J_z+A_z}$ & 5\\
Double Core& $\checkmark$ & $\checkmark$ & $\checkmark$ & $\checkmark$ & $(S^3/\mathbb{Z}^2)_{J} \times S^1_A \times S^2_t$ & 6 \\
\br
\end{tabular}
\item[] $^{\rm a}$ We have noted the equivalence of $\delta$ and $\omega_z$ moduli for the A-phase core.
\end{indented}
\end{table}

\section{Effective Dynamics of Vortex Moduli Fields}

In the previous section we considered the symmetries of the free energy (\ref{GLFE}) to determine the possible moduli fields appearing on the effective theory of vortex excitations in the B-phase.  In this section we will determine the effective field theory for the specific vortex solutions (\ref{ResDec}), (\ref{OriginalAnsatz}), (\ref{ComplexAnsatz}), and (\ref{DoubleCoreAnsatz}) considered above.  We will begin by outlining the general procedure for determining the low energy effective theory of vortex fluctuations (see \cite{Shifman:2012zz} and \cite{Eto:2005yh}).  For this purpose we will follow the construction outlined in \cite{Nitta:2013mj}.

In general, the effective field theory can be determined by considering fluctuations of the vortex solution
\begin{equation}
e_{\mu i}(\vec{x}_{\perp}) = e^{\rm vort}_{\mu i}(\vec{x}_{\perp})+\delta e_{\mu i}(x,y,z,t).
\label{Fluctuations}
\end{equation}
We insert (\ref{Fluctuations}) into the free energy (\ref{GLFE}) and consider terms of second order in $\delta e_{\mu i}$.  Integrating by parts in the spatial gradient terms we arrive at the following equation
\begin{eqnarray}
&\delta^2 F_{GL} = i \delta e_{\mu i} \partial_t \delta e_{\mu i}^{\star}+\delta e_{\mu i} L_{ij,\mu\nu}\delta e_{\nu j}^{\star}, \nonumber \\
& L_{ij,\mu\nu} = -\gamma_1\delta_{ij}\delta_{\mu\nu}\vec{\partial}^2-\gamma_{23}\delta_{\mu\nu}\partial_i \partial_j+(\partial_{e_{\mu i}^{\star} }\partial_{e_{\nu j}}V).
\label{EFT}
\end{eqnarray}
Here we have consolidated the spatial gradient and potential terms to the operator $L_{ij,\mu\nu}$.  For our present purposes it will only be necessary to consider $\delta^2F_{GL}$ up to second order in $\delta e_{\mu i}$.  Allowing $\delta e_{\mu i}$ to depend only on $\vec{x}_{\perp}$ for the moment and varying (\ref{EFT}) with respect to $\delta e_{\mu i}$ we arrive at an eigenvalue equation
\begin{equation}
L_{i j}(\vec{x}_{\perp})e_{\mu j}^{(n)}(\vec{x}_{\perp})=E^{(n)}e_{\mu i}^{(n)}(\vec{x}_{\perp}),
\end{equation}
where the fluctuations $\delta e_{\mu i}$ are written in terms of the eigenmodes $e_{\mu i}^{(n)}$
\begin{equation}
\delta e_{\mu i}(x,y,z,t) = \sum_n c_n(t,z) e_{\mu i}^{(n)}(x,y).
\end{equation}
Here we have restored the $t$ and $z$ dependences in $\delta e_{\mu i}$ by assuming the adiabatic approximation.
The gapless modes are determined by the eigenmodes with eigenvalues $E^{(n)} = 0$.  They are defined by transformations of the vortex solution $e_{\mu i}^{\rm vort}$ that leave the Ginzburg-Landau free energy (\ref{GLFE}) invariant.  The set of moduli defining these symmetry transformations are denoted by $m^a$ and we may write the gapless fluctuations as
\begin{equation}
\delta e_{\mu i}(x,y,z,t) =\sum_a m^a(t,z)\frac{\partial}{\partial m^a}e_{\mu i}^{\rm vort}(x,y,m).
\label{GaplessFluctuations}
\end{equation}
Inserting (\ref{GaplessFluctuations}) into the free energy (\ref{EFT}) and ignoring all other modes with non-zero energy $E^{(n)}$ and integrating over $x$ and $y$ we arrive at the effective field theory of $m^a(t,z)$ describing the string like dynamics of the vortex solution.  If the moduli we consider are strictly gapless we will find the free energy can be written as
\begin{eqnarray}
&F_{\rm eff}=iG_{ab}(m)m^a \partial_t m^b+G_{ab}(m)\partial_z m^a \partial_z m^b, \nonumber \\
&G_{ab} = \int d^2 \vec{x}_{\perp}\frac{\partial e_{\mu i}^{\rm vort}(\vec{x}_{\perp},m)}{\partial m^a} \frac{\partial e_{\nu j}^{\rm vort}(\vec{x}_{\perp},m)}{\partial m^b} 
\end{eqnarray}
where $G_{ab}$ is a function of the moduli fields $m^a$, and is symmetric in $a$ and $b$.

For the particular problem under consideration the moduli fields $m^a$ are given by (\ref{ModuliList}), however some of these moduli may not appear in the effective free energy if their generators annihilate the vortex solution.

When we switch on $\gamma_2,\gamma_3 \neq 0$, the number of gapless moduli appearing in (\ref{ModuliList}) will be restricted to (\ref{RefinedModuliList}) and the remaining moduli will acquire either a mass gap proportional to $\gamma_{23}$ or acquire interactions with the gapless moduli from (\ref{RefinedModuliList}).  We will expect these interactions to occur for the case of $\xi_{x,y}$ and $\omega_{x,y}$. Only a combination of these moduli fields will be strictly gapless.  This is simply a reflection of the locking of coordinate and orbital rotations, and the equivalence of coordinate rotations with $z$-dependent translations when $\gamma_2,\gamma_3$ are non-zero.  If we assume $\gamma_2,\gamma_3$ are small compared to $\gamma_1$ we may consider the additional moduli as quasi-gapless.  

We proceed by first considering the ansatz (\ref{ResDec}) with $\chi^k$ field given by (\ref{OriginalAnsatz}).  We assume for the moment that $\gamma_{23} = 0$ and thus the translational moduli $\xi_a$ and rotational moduli $\omega_a$ decouple.  The moduli appear as the following transformations
\begin{eqnarray}
&e_{\mu i}(\vec{x}_{\perp}) \rightarrow e_{\mu i}(\vec{x}_{\perp}-\vec{\xi}_{\perp}), \; \; \xi \in t \nonumber \\
&\chi^i(r_{\perp}) \rightarrow R_{ij}(\vec{\omega})\chi^j(r_{\perp}), \; \;R_{ij} \in SO(3)_{S+L} \nonumber \\
&\chi^i(r_{\perp}) \rightarrow e^{i \delta}\chi^i(r_{\perp}), \; \;e^{i \delta} \in U(1)_A,
\label{ModuliTrans}
\end{eqnarray}
where $\vec{\xi}_{\perp}$, $\vec{\omega}$, and $\delta$ are functions of $z$ and $t$.  Additionally, it will be particularly convenient to consider the rotational moduli in the form
\begin{equation}
\chi^i = R_{ij}(\vec{\omega})\chi^j \equiv S^i(t,z)\chi(r_{\perp}), \; \; |S|^2  = 1,
\end{equation}
and consider the real moduli fields $\vec{S}(t,z)$ instead of $\vec{\omega}(t,z)$.  We are permitted to make this switch of variables because the particular vortex solution (\ref{OriginalAnsatz}) retains an orbital $SO(2)$ symmetry and thus the solution has only two moduli fields from the degeneracy space $SO(3)_{S+L}/SO(2)$, which define the unit vector $S^k(t,z)$ in the direction of $\chi^k(r_{\perp},z,t)$.

Performing the general procedure outlined above we may immediately write out the effective theory of the moduli fields
\begin{eqnarray}
F_{\rm eff} = F_{\rm trans}+F_{O(3)}+F_{U(1)}, \nonumber \\
F_{\rm trans}=\frac{T}{2}\partial_z \vec{\xi}_{\perp}\partial_z \vec{\xi}_{\perp}, \nonumber \\
F_{O(3)} = \frac{1}{2g^2}\partial_z \vec{S} \partial_z \vec{S}, \; |S|^2 \equiv 1, \nonumber \\
F_{U(1)} = \frac{1}{2g^2}\partial_z \delta \partial_z \delta,
\label{EFTOriginal}
\end{eqnarray}
where we are omitting the time derivatives.  We note that strictly speaking the field $\rho$ cannot be dynamical, since it's time derivative term in the free energy could be written as a total derivative.  

The couplings $T$ and $g^2$ are determined from the integration over $\vec{x}_{\perp}$:
\begin{eqnarray}
\frac{T}{2} \sim \int d^2\vec{x}_{\perp} 3\gamma_1\frac{f^2}{r_{\perp}^2} \rightarrow \gamma_1\Delta^2 \log \left(\frac{\alpha R^2}{\gamma_1}\right), \\
\frac{1}{g^2} \sim  \int d^2\vec{x}_{\perp}\gamma_1\chi^2 \rightarrow  \frac{\gamma_1^2}{2\beta_{12}+\beta_{345}},
\label{Feff}
\end{eqnarray}

As expected $\vec{S}$ and $\vec{\xi}_{\perp}$ decouple, and the effective theory takes the form of an $O(3)$ sigma model in one spatial dimension, plus a translational part that describes the Kelvin excitations, and a $U(1)_A$ part describing axial rotations.  We see that the theory (\ref{EFTOriginal}) has a total of five gapless moduli fields.

At this point we switch on a small but non-zero $\gamma_{23}$.  We will assume that $\gamma_{23}$ is small enough that we may neglect the corrections to the vortex solutions of $f(r_{\perp})$ and $\chi(r_{\perp})$, as well as the constants $T$ and $g^2$.  Aside from these uninteresting numerical corrections (\ref{Feff}) remains of the same form, however there are additional terms representing the breaking of the $SO(3)_{S+L} \times SO(3)_{\rm coord}$ symmetries.  They appear as follows
\begin{equation}
F_{\rm eff} \rightarrow F_{\rm eff}-\Delta F_{\rm eff},
\end{equation}
where $\Delta F_{\rm eff}$ represents symmetry breaking terms, which in this case are given by
\begin{eqnarray}
\Delta F_{\rm eff} = M^2(\vec{S}_{\perp}-S^3\partial_z \vec{\xi}_{\perp})^2 +\frac{\varepsilon}{2g^2}\left\{(S^3 \partial_z \delta)^2+(\partial_z S^3)^2\right\}.
\label{ModFeff}
\end{eqnarray}
Here $\varepsilon \sim \gamma_{23}/\gamma_1$, and $M$ represents a mass gap parameter given by
\begin{equation}
M^2 \sim \int d^2\vec{x}_{\perp}\gamma_{23} (\partial_{\perp} \chi)^2 \rightarrow \frac{\gamma_{23}\alpha}{2(2\beta_{12}+\beta_{345})}.
\end{equation}
In this form $M$ is known as the ``twisted mass" \cite{Alvarez:1983, Gates:1984}.

If we expand the first term in (\ref{ModFeff}) we find that the mass gap $M^2$ represents a negative mass term (if $\gamma_{23} > 0$) for $\vec{S}_{\perp}$ and implies a minimum energy for $|S_{\perp}|^2 = 1$, and thus $S^3 \rightarrow 0$.  This characterizes the explicit $O(3)$ symmetry breaking in which the lowest energy state retains a $U(1)$ degeneracy.  If $\gamma_{23} < 0$ the mass term will be positive and we will find $\vec{S}_{\perp} \rightarrow 0$, in which case the $O(3)$ is completely broken with no degeneracy of the ground state.  We also note that the first term of (\ref{ModFeff}) implies a correlation between the translational moduli $\vec{\xi}_{\perp}$ and the rotational moduli $\vec{S}_{\perp}$.  If $S^3 \neq 0$ we find the free energy is minimized for $\partial_z \vec{\xi}_{\perp}$ antiparallel to $\vec{S}_{\perp}$.  This is of course to be expected due to the locking of the coordinate and orbital rotations when $\gamma_{23} \neq 0$.

We point out the agreement of the results (\ref{EFTOriginal}) and (\ref{ModFeff}) with the low energy effective dynamics of moduli fields emerging on the ANO string (see \cite{Shifman:2013oia} and \cite{Monin:2013kza}).  Note that since we consider a global $U(1)_P$ phase symmetry we have the additional axial modulus $\delta(z,t)$ appearing in the effective theory (\ref{EFTOriginal}).

We may extend this analysis to the anti-symmetric A-phase ansatz (\ref{ComplexAnsatz}).  Here the translational moduli $\vec{\xi}_{\perp}(z)$ are generated as in the previous example in (\ref{ModuliTrans}).  The $SO(3)_{S+L}$ and $U(1)_A$ moduli are generated by
\begin{equation}
\vec{\chi}_A(r_{\perp},\phi) \rightarrow e^{i \delta} \chi_0(r_{\perp})\vec{S}+e^{-i \delta}\chi_2(r_{\perp})\vec{S}^{\star}e^{2i\phi},
\label{ComplexModTrans}
\end{equation}
where $\delta$ and $\vec{S}$ are functions of $z$.  Here the unit vector field $\vec{S}$ has been promoted to a complex vector 
\begin{equation}
\vec{S} \equiv (\vec{S}_R + i\vec{S}_I)/\sqrt{2},
\end{equation}
with the following conditions:
\begin{equation}
|\vec{S}_R| = |\vec{S}_I|=1, \mbox{ and }  \vec{S}_R \cdot \vec{S}_I = 0.
\end{equation}  

At this point we perform the same procedure as the previous example to determine the form of the effective free energy describing the moduli dynamics.  We again write $F_{\rm eff}$ as a combination of $\gamma_1$ and symmetry breaking $\gamma_{23}$ terms.  For $\gamma_{23} = 0$ we have
\begin{eqnarray}
&F_{\rm eff} = F_{\rm trans}+F_{O(3)+U(1)}, \nonumber \\ 
&F_{\rm trans}=\frac{T}{2}(\partial_z \vec{\xi}_{\perp})^2, \nonumber \\ 
&F_{O(3)+U(1)} =\frac{1}{2g^2}|\partial_z \vec{S}+i\vec{S}\partial_z\delta|^2,
\label{ModEFT}
\end{eqnarray}
where $T$ is similar to the case in (\ref{Feff}).  However, $g^2$ is now given by
\begin{equation}
\frac{1}{2g^2} \sim  \int d^2\vec{x}_{\perp}\gamma_1\chi_{0,2}^2 \rightarrow \frac{\gamma_1^2}{\alpha}\Delta^2\log\left(\frac{\alpha R^2}{\gamma_1}\right),
\end{equation}
due to the $1/r_{\perp}$ nature of $\chi(r_{\perp})$ as $r_{\perp} \rightarrow \infty$.

We have written $F_{O(3)+U(1)}$ in (\ref{ModEFT}) in a form emphasizing the symmetry (\ref{HiddenSymm}).  Thus, although the complex nature of $\vec{S}$ implies a complete breaking of $SO(3)_{S+L}$ and hence three complete non-Abelian moduli, the hidden symmetry (\ref{HiddenSymm}) reduces the total number of moduli by one.  In this case we have a total of five gapless moduli.

We continue by including the corrections when $\gamma_{23} \neq 0$.  Writing 
\begin{equation}
F_{\rm eff} \rightarrow F_{\rm eff}-\Delta F_{\rm eff},
\end{equation}
with
\begin{equation}
\Delta F_{\rm eff} = M^2|\vec{S}_{\perp} -S^3\partial_z \vec{\xi}_{\perp}|^2 + \frac{\varepsilon}{2g^2}|\partial_z S^3 + i S^3 \partial_z \delta|^2.
\label{ModEffAphase}
\end{equation}
Here $\varepsilon = \gamma_{23}/\gamma_1$ as previously and $M^2$ is given by
\begin{equation}
M^2 \sim \int d^2\vec{x}_{\perp}\gamma_{23} (\partial_{\perp} \chi_{0,2})^2 \rightarrow \gamma_{23} \Delta^2\times C(\beta_i),
\end{equation}
where $C(\beta_i)$ is a constant dependent on the $\beta_i$ coefficients.  We see again that (\ref{ModEffAphase}) shows the correlation between the orbital $SO(3)_{S+L}$ rotations and $z$-dependent translations due to the breaking of $SO(3)_L \times SO(3)_{\rm coord} \rightarrow SO(3)_{L+R} \equiv SO(3)_L$ by the non-zero $\gamma_2,\gamma_3$ in the gradient terms of (\ref{GLFE}).

We complete this analysis by considering the moduli fields of the more general double core vortex ansatz (\ref{DoubleCoreAnsatz}).  Again, the translational moduli $\vec{\xi}_{\perp}(z)$ are generated by the transformation in (\ref{ModuliTrans}).  We represent the $SO(3)_{S+L}$ and $U(1)_A$ moduli by
\begin{eqnarray}
\fl \vec{\chi}_{dc}(r_{\perp},\phi) \rightarrow \frac{1}{\sqrt{2}}(e^{i \delta} \chi_0(r_{\perp})+e^{-i \delta}\chi_2(r_{\perp})e^{2i\phi})\vec{S} \nonumber \\
+ \frac{i}{\sqrt{2}}(e^{i \delta} \zeta_0(r_{\perp})-e^{-i \delta}\zeta_2(r_{\perp})e^{2i\phi})\vec{n},
\label{DoubleCoreModTrans}
\end{eqnarray}
where $\vec{S}$ and $\vec{n}$ are both real unit vectors, which are orthogonal to each other
\begin{equation}
\vec{S} \cdot \vec{n} \equiv 0.
\end{equation}
For $\gamma_{23} = 0$ we write the effective free energy following the same procedure as previously
\begin{eqnarray}
\fl F_{\rm eff} =  \frac{T}{2}(\partial_z \vec{\xi}_{\perp})^2+\frac{1}{2 g_{\delta}^2}(\partial_z \delta)^2 +\frac{1}{2g_{\chi}^2}(\partial_z \vec{S})^2+\frac{1}{2g_{\zeta}^2}(\partial_z \vec{n})^2 \nonumber \\
+\frac{1}{g_{\chi \zeta}^2}(\partial_z \delta)(\vec{S} \cdot \partial_z \vec{n}-\vec{n} \cdot \partial_z \vec{S}),
\label{EFTdc}
\end{eqnarray}
where the tension $T$ is similar in value to the previous examples, and the couplings $g_{\chi}, \; g_{\zeta}, \; g_{\chi\zeta}$, and $g_{\delta}$ are given by
\begin{eqnarray}
&\frac{1}{2 g_{\chi}^2} =\frac{1}{2}\int d^2 \vec{x}_{\perp}\gamma_1 \chi_{0,2}^2 \rightarrow \frac{\gamma_1^2}{\alpha}\Delta^2 \log\left(\frac{\alpha R^2}{\gamma_1}\right), \nonumber \\ 
&\frac{1}{2 g_{\zeta}^2} = \frac{1}{2}\int d^2 \vec{x}_{\perp}\gamma_1 \zeta_{0,2}^2 \sim \frac{1}{2g_{\chi}^2}, \nonumber \\ 
&\frac{1}{2g_{\chi\zeta}^2} = \frac{1}{2}\int d^2 \vec{x}_{\perp}\gamma_1 (\chi \zeta)_{0,2} \sim \frac{1}{2g_{\chi}^2}, \nonumber \\ 
&\frac{1}{2g_{\delta}^2} = \frac{1}{2 g_{\chi}^2}+\frac{1}{2 g_{\zeta}^2}.
\end{eqnarray}
As long as $1/g_{\chi\zeta}^2 \neq 1/g_{\chi} g_{\zeta}$ there is no hidden symmetry, and we indeed find a total of six independent moduli from the complete breaking of $SO(3)_{S+L} \times U(1)_A \times t$ by the double core vortex.  

Turning on $\gamma_{23}$ we would again find correlations between $(\vec{S},\vec{n})$, and  $\vec{\xi}_{\perp}$ as well as mass terms for $S^3$ and $n^3$, all proportional to $\gamma_{23}$.  The complete form for $\Delta F_{\rm eff}$ for the double core vortex is however complex and we omit it here.

\section{Discussion and Conclusions}
In the analysis considered here we have explored the Ginzburg-Landau description of superfluid $^3$He and similar systems with tensorial order parameter and non-Abelian group structure.  The symmetry structure of such systems allows for the investigation of non-Abelian gapless and quasi-gapless moduli localized on mass vortices.  We have attempted to illustrate these concepts by applying techniques standard to high energy physics to condensed matter systems like superfluid $^3$He-B.  In particular, we have given a general procedure for determining effective theory describing the low energy excitations of vortices in superfluid $^3$He.

Additionally we have applied this procedure to a specific case (as suggested in \cite{Nitta:2013mj}) and derived the low energy effective theories for the gapless and quasi-gapless excitations of topological vortices appearing in the B-phase of superfluid $^3$He.  This was accomplished by considering the ansatz for the order parameter given by (\ref{ResDec}), with anti-symmetric components given by (\ref{OriginalAnsatz}) and (\ref{ComplexAnsatz}).  We have chosen this specific ansatz because it illustrates the process of determining the non-Abelian moduli fields without introducing large amounts of calculation.  However, it is well known both theoretically \cite{Thuneberg:1986a, Volovik:1986a} and experimentally \cite{Krusius:1984a, Pekola:1985a, Hakonen:1983a} that the minimizing solutions contain additional terms in the order parameter.  Considering more complex vortex solutions would not change the moduli involved in the low energy theory, however their interactions and coefficients may change depending on the symmetries respected by the vortex solution.

The low energy effective field theories we have derived given by (\ref{EFT}) and (\ref{ModEFT}) with the respective $\gamma_{23}$ corrections (\ref{ModFeff}) and (\ref{ModEffAphase}), exhibit the form of one-dimensional $O(3)$ sigma models whose moduli fields interact with translational moduli generating the well known Kelvin modes  \cite{Thomson:1880, Sonin:1987zz, Simula:2008a, Fonda,Krusius:1984a, Pekola:1985a, Hakonen:1983a}.  In particular, the model (\ref{EFT}) with (\ref{ModFeff}), is very similar to the $1+1$-dimensional model describing the low energy dynamics of ANO strings in Yang-Mills theories \cite{Abrikosov:1957, Nielsen:1973}.  We do however point out that the present theory has the additional $U(1)$ modulus $\delta$ appearing due to the global phase symmetry $U(1)_P$ of (\ref{GLFE}).  In the Yang-Mills theories describing the ANO strings, the $U(1)$ modulus $\delta$ can be removed by a corresponding gauge transformation.  This is not possible in the present situation and thus the $U(1)$ modulus $\delta$ is unavoidable.  It is however emphasized that the modulus $\delta$ does not propagate since its corresponding time derivative term in the free energy (\ref{GLFE}) reduces to a total derivative.

We wish to emphasize that our analysis of the effective field theory is purely classical, and we have made no attempt to discuss quantization.  It is expected that after quantization the number gapless excitations appearing in the effective theory may be different from the number of moduli fields.  This is of course well known for the case of the Kelvin mode, in which two moduli describing translational excitations, actually imply only a single mode after quantization.  This is a manifestation of the Goldstone theorem applied to non-relativistic systems in which the number of modes may be equal to or less than the number of broken symmetries \cite{Nielsen:1976, Watanabe:2012, Hidaka:2012ym}.  We expect that such a reduction of non-Abelian modes may appear in a similar fashion to the Kelvin modes.  Future analysis will be devoted to this investigation.

\ack
The authors would like to thank G. Volovik for his comments and suggestions.  A. P. would like to thank J. Kapusta for interesting discussions.  M. S. is grateful to Z. Komargodski for his comments.  This work was supported in part by DOE grant DE-FG02-94ER40823.

\end{document}